\documentclass[twocolumn]{aastex63}

\usepackage{amsmath}
\usepackage{algorithmic}

\received{2020 July 28}
\revised{2020 September 15}
\accepted{2020 September 17}

\submitjournal{ApJ}

\shorttitle{ICME rate for solar cycle 25 and implications for Parker Solar Probe}
\shortauthors{M\"ostl et al. }
\begin{document}

\title{Prediction of the in situ coronal mass ejection rate for solar cycle 25: Implications for Parker Solar Probe in situ observations}


\correspondingauthor{Christian M\"ostl}
\email{christian.moestl@oeaw.ac.at}

\author[0000-0001-6868-4152]{Christian M\"ostl}
\affiliation{Space Research Institute, Austrian Academy of Sciences, Schmiedlstraße 6, 8042 Graz, Austria}
\affiliation{Institute of Geodesy, Graz University of Technology, Steyrergasse 30, 8010 Graz, Austria}

\author[0000-0002-6273-4320]{Andreas J. Weiss}
\affiliation{Space Research Institute, Austrian Academy of Sciences, Schmiedlstraße 6, 8042 Graz, Austria}
\affiliation{Institute of Geodesy, Graz University of Technology, Steyrergasse 30, 8010 Graz, Austria}
\affiliation{Institute of Physics, University of Graz, Universit\"atsplatz 5, 8010 Graz, Austria}

\author[0000-0003-2021-6557]{Rachel L. Bailey}
\affiliation{Space Research Institute, Austrian Academy of Sciences, Schmiedlstraße 6, 8042 Graz, Austria}
\affil{Zentralanstalt f\"ur Meteorologie und Geodynamik, Hohe Warte 38, 1190 Vienna, Austria}

\author[0000-0002-6362-5054]{Martin A. Reiss}
\affiliation{Space Research Institute, Austrian Academy of Sciences, Schmiedlstraße 6, 8042 Graz, Austria}
\affiliation{Institute of Geodesy, Graz University of Technology, Steyrergasse 30, 8010 Graz, Austria}

\author[0000-0001-9024-6706]{Tanja Amerstorfer}
\affiliation{Space Research Institute, Austrian Academy of Sciences, Schmiedlstraße 6, 8042 Graz, Austria}

\author[0000-0002-1222-8243]{J\"urgen Hinterreiter}
\affiliation{Space Research Institute, Austrian Academy of Sciences, Schmiedlstraße 6, 8042 Graz, Austria}
\affil{Institute of Physics, University of Graz, Universit\"atsplatz 5, 8010 Graz, Austria}

\author[0000-0002-2507-7616]{Maike Bauer}
\affiliation{Space Research Institute, Austrian Academy of Sciences, Schmiedlstraße 6, 8042 Graz, Austria}
\affil{Institute of Physics, University of Graz, Universit\"atsplatz 5, 8010 Graz, Austria}

\author[0000-0002-7369-1776]{Scott W. McIntosh}
\affiliation{National Center for Atmospheric Research, P.O. Box 3000, Boulder, CO 80307, USA}

\author[0000-0002-1890-6156]{No\'e Lugaz}
\affiliation{Space Science Center, Institute for the Study of Earth, Oceans, and Space, University of New Hampshire, Durham, NH, USA}

\author[0000-0002-1365-1908]{David Stansby}
\affiliation{Mullard Space Science Laboratory, University College London, Holmbury St. Mary, Surrey RH5 6NT, UK}

\begin{abstract}
The Parker Solar Probe (PSP) and Solar Orbiter missions are designed to make groundbreaking observations of the Sun and interplanetary space within this decade. We show that a particularly interesting in situ observation of an interplanetary coronal mass ejection (ICME) by PSP may arise during close solar flybys ($< 0.1$~AU). During these times, the same magnetic flux rope inside an ICME could be observed in situ by PSP twice, by impacting its frontal part as well as its leg. Investigating the odds of this situation, we forecast the ICME rate in solar cycle 25 based on 2 models for the sunspot number (SSN): (1) the forecast of an expert panel in 2019 (maximum SSN = 115), and (2) a prediction by McIntosh et al. (2020, maximum SSN = 232). We link the SSN to the observed ICME rates in solar cycles 23 and 24 with the Richardson and Cane list and our own ICME catalog, and calculate that between 1 and 7 ICMEs will be observed by PSP at heliocentric distances $< 0.1$ AU until 2025, including 1$\sigma$ uncertainties. We then model the potential flux rope signatures of such a double-crossing event with the semi-empirical 3DCORE flux rope model, showing a telltale elevation of the radial magnetic field component $B_R$, and a sign reversal in the component $B_N$ normal to the solar equator compared to field rotation in the first encounter. This holds considerable promise to determine the structure of CMEs close to their origin in the solar corona.
\end{abstract}



\section{Introduction} \label{sec:intro}

Coronal mass ejections (CMEs) form a major link between a star and planetary bodies within the astrosphere. In the solar system, the knowledge of how many CMEs erupt during the solar cycle and how often we see their manifestations at spacecraft observing the in situ solar wind are of great interest in order to assess the potential of CMEs as drivers of major geomagnetic storms \citep{Richardson2000, zhang2007}. If a spacecraft observes a CME in situ, we call the full interval of disturbed solar wind an interplanetary coronal mass ejection \citep[ICMEs, for an exact definition see][]{rouillard2011}.

A major result of previous studies is that ICMEs drive all major geomagnetic storms with Dst~$< - 250$~nT, with the Disturbance storm time (Dst) index as a measure of the severity of geomagnetic storms at Earth. \citet{love2015_historic} found that a geomagnetic disturbance with a Dst of roughly $ - 1000$~nT, generally accepted to be the order of magnitude of the Carrington event \citep{carrington1859} and the May 1921 storm \citep{love2019_may21}, happens once in 100 years. Given the average ICME rate at Earth \citep[e.g.][]{Richardson2000} with about 20 events per year, we can easily see that in $10^2$ years about $2 \times 10^3$ ICMEs impact the Earth, out of which a single one is expected to elicit a Carrington magnitude geomagnetic storm. CMEs are now widely considered to be a natural hazard, as the probability of a Carrington-type event is $10\%$ per solar cycle and knowing the ICME rate at Earth and other solar system locations is a basic need in order to study, forecast and mitigate their effects.

The idea for this paper originated from simulations that we have done with our semi-empirical 3DCORE model which can describe the magnetic flux ropes inside ICMEs \citep{moestl_2018}. When the Parker Solar Probe \citep[PSP,][]{fox2016_psp} spacecraft passes close to the Sun in the upcoming years, an ICME flux rope could be crossed twice by PSP as it moves extremely fast during those solar flybys. We have thus wondered how often this situation might happen during the PSP primary mission until 2025, and thus considered to forecast the ICME rate for the next solar cycle to calculate how many such events we could expect. This paper is thus essentially divided into 2 parts: the first where we predict the ICME rate for solar cycle 25, and the second where we model a CME encounter during a PSP solar flyby in an idealized way.

\citet{Richardson2001}, \citet{Jian2006} and other works by the same authors have clarified many questions on the occurrence rate of ICMEs \citep[for a summary see][]{webb2012_review}. However, during a solar cycle the rate of ICMEs impacting different points in the heliosphere can vary considerably from location to location. Using accurate statistics to forecast the occurrence rate of ICMEs as potentially measured by PSP in the innermost heliosphere and upper corona is critical. In a previous study, we compiled the ICMECAT catalog \citep{Moestl2017}. It is an aggregate of different individual catalogs along with our own identified ICMEs from data provided by the missions STEREO, Wind, Venus Express, Ulysses, and MESSENGER. The catalog contains data from 2007 onwards and was originally constructed to validate CME prediction models at various targets within the inner heliosphere. We present here an updated ICMECAT to include ICMEs at the aforementioned missions until the end of 2019, have added events observed by the MAVEN spacecraft at times when it sampled the solar wind.


In \textbf{section \ref{sec:methods}}, we show that our combined ICMECAT catalog allows an assessment of the ICME rate better than previous studies, as multi-point locations give a better ICME sample size. We can thus determine the random spread in ICME numbers per year at different in situ locations directly from data. The ICME rate is a stochastic phenomenon, as the Sun does not produce CMEs in a steady way but CMEs erupt from the Sun at a high rate during the presence of active regions, followed by times with no or few CMEs, or erupt with slow speeds from regions of the otherwise quiet Sun. The ICME rate does not depend on heliocentric distance, as it is well established that CMEs expand self-similarly \citep{yashiro2004}, which means they retain the same angular width as they propagate away from the Sun.

\textbf{Section \ref{sec:results}} estimates the number of ICMEs observed at Earth and other heliospheric locations including the Parker Solar Probe, Solar Orbiter \citep{Mueller_2013} and Bepi Colombo up until 2030. To this end, we combine the relationship that we derived between the ICME rate and the Sunspot number with two different predictions for the magnitude of solar cycle 25, assuming one of these solar cycle predictions holds roughly true. Additionally, the PSP mission, launched in August 2018, will perform many close approaches to the Sun as close as 0.05 AU, which will give us an unprecedented opportunity to observe an ICME flux rope with high relative tangential speed between the spacecraft and ICME. PSP will attain speeds of up to 180 km~s$^{-1}$, which is close to the lower end of the speed distribution of CMEs. We show that this observing situation may lead to a double-crossing of the same ICME flux rope by PSP.  

We further investigate the expected magnetic signatures of such an observation, aided by modeling with the 3DCORE semi-empirical flux rope technique. The 3DCORE method allows both forward modeling and fitting of 3D magnetic field flux rope signatures in retrospect. It is based on a global torus shape with elliptical cross section, a Gold Hoyle field \citep{gold_hoyle_1960} and drag-based kinematics \citep{vrsnak_2013}. Here, we use the new version of \citet{weiss2020} with a highly streamlined numerical calculation pipeline, considerably improving the  \citet{moestl_2018} 3DCORE prototype. 

\section{Methods} \label{sec:methods}

We start by describing our ICME catalog, and proceed to establish a correlation with the sunspot number as a proxy for solar activity, allowing us to derive a forecast of the ICME rate for the next solar cycle.

\subsection{ICME catalog}

\begin{figure*}[ht!]
\noindent\includegraphics[width=\linewidth]{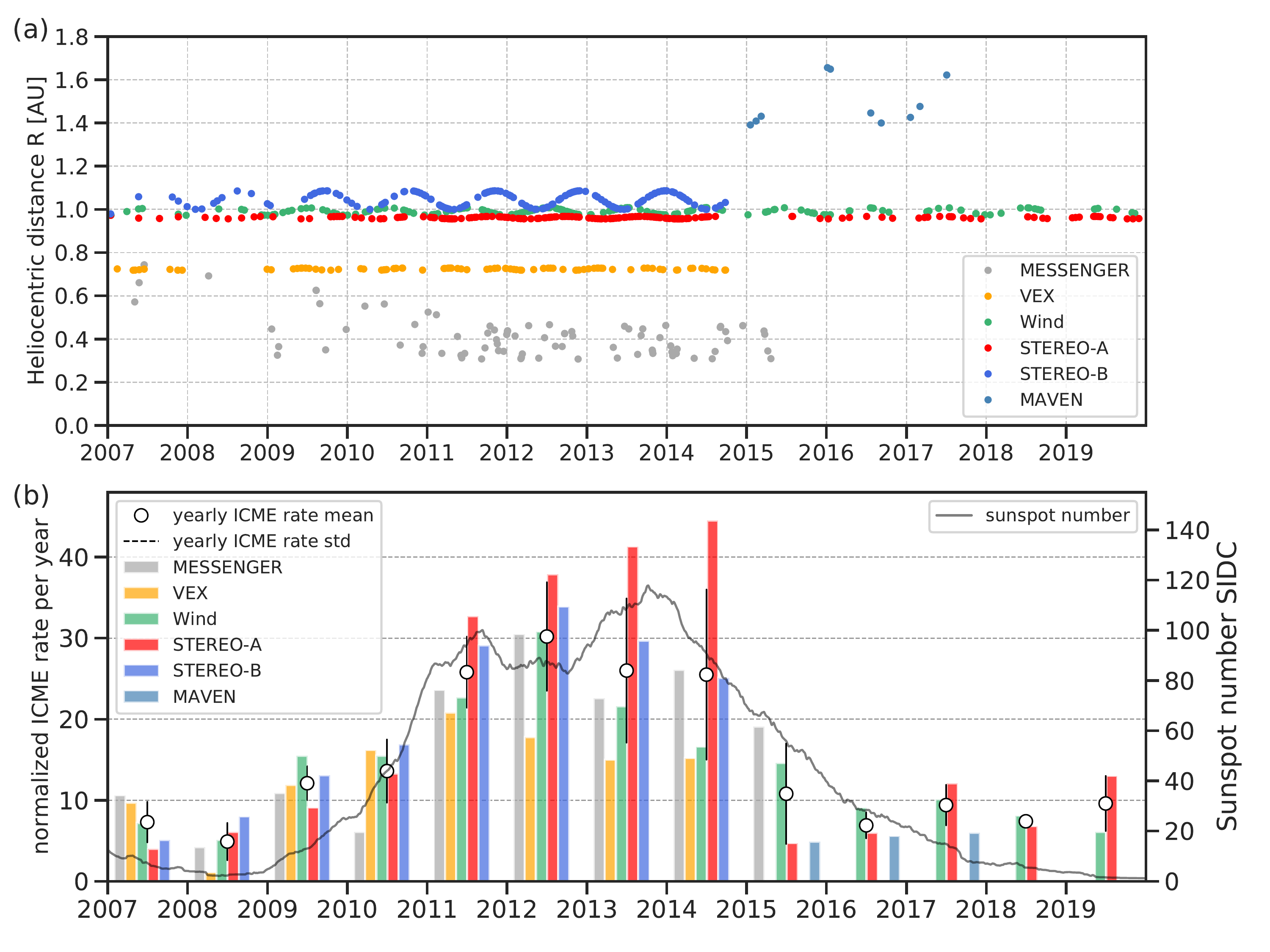}
\caption{ICMECAT data coverage and impact frequency. (a) ICME event detections as a function of time and heliocentric distance. The spacecraft is given by the color code in the legend. (b) Number of ICMECAT events per year for each spacecraft, corrected for in situ observation gaps. White dots indicate the mean for each year with the error bars as the standard deviation. The sunspot number from SIDC is plotted for orientation with respect to the solar cycle progress.}
\label{fig:frequency}
\end{figure*}

First, we present the most comprehensive catalog of interplanetary coronal mass ejections to date, including a total of 739 events at MESSENGER in the cruise phase and in orbit around Mercury, at VEX in orbit around Venus, at Wind at the Sun-Earth L1 point, at STEREO-A/B in the solar wind near 1 AU, at Ulysses during its last ecliptic pass in 2007, and at MAVEN in orbit around Mars since September 2014. Our update here extends the ICMECAT catalog first shown by \cite{Moestl2017}, which originally contained 668 events. The time range now spans 1 January 2007 -  31 December 2019. 

The ICME times were taken from various authors \citep[see][]{Moestl2017} or newly included in the lists by ourselves if we had data intervals for which no ICME categorization had been performed by other authors to date (valid for MESSENGER, VEX, Ulysses, and MAVEN). All ICME parameters were then consistently recalculated from the data. The ICMECAT contains parameters for each event based on magnetic field and bulk plasma observations, when available. From the up to 30 parameters determined for each event, in this study we use only the existence of the event and its ICME start time. Further details on how the ICMECAT was established are described in \cite{Moestl2017}. The most recent catalog version is available on our webpage\footnote{\url{https://helioforecast.space/icmecat}}, and an archived release of the catalog used in this paper, in many different formats and including a citable DOI, can be found on figshare (see data sources in the acknowledgements).

\textbf{Figure \ref{fig:frequency}a} shows the ICME event coverage with each dot being an ICME observed at the indicated time and heliocentric distance, with a color code distinguishing each spacecraft. We use Wind data for the continuously observed near Earth space at the Sun--Earth L1 point. The plot further shows a decent general coverage of the inner heliosphere $< 1$~AU during solar cycle 24, with more sparse event detections towards the end of the cycle as the VEX and MESSENGER missions ended in 2014 and 2015, respectively.

\textbf{Figure \ref{fig:frequency}b} demonstrates the number of ICMECAT events per year for each spacecraft. For each data point, the number of ICMEs per year was normalized according to the data availability for the given year, thus longer data gaps were identified in the magnetic field data and excluded in the calculation. For years with little data availability, the histogram entry was set to NaN. The plot shows a clear rise of the yearly ICME rate (ICR) from solar minimum to maximum, as well as a decrease in the declining phase to the current very low rates during solar minimum. The number of ICMEs has risen from an average of 5 events during solar minimum to  around 30 events per year in solar maximum, or about 2.5 events per month. The spread in the ICME rate at different observatories is given as an error bar on \textbf{Figure \ref{fig:frequency}b}. The standard variation in the ICME rate from including all observatories varied from about $\pm 2$ ICMEs in solar minimum to $\pm 10$ events in solar maximum.

\subsection{Correlation of ICME rate and sunspot number}

\begin{figure}[t]
\noindent\includegraphics[width=\linewidth]{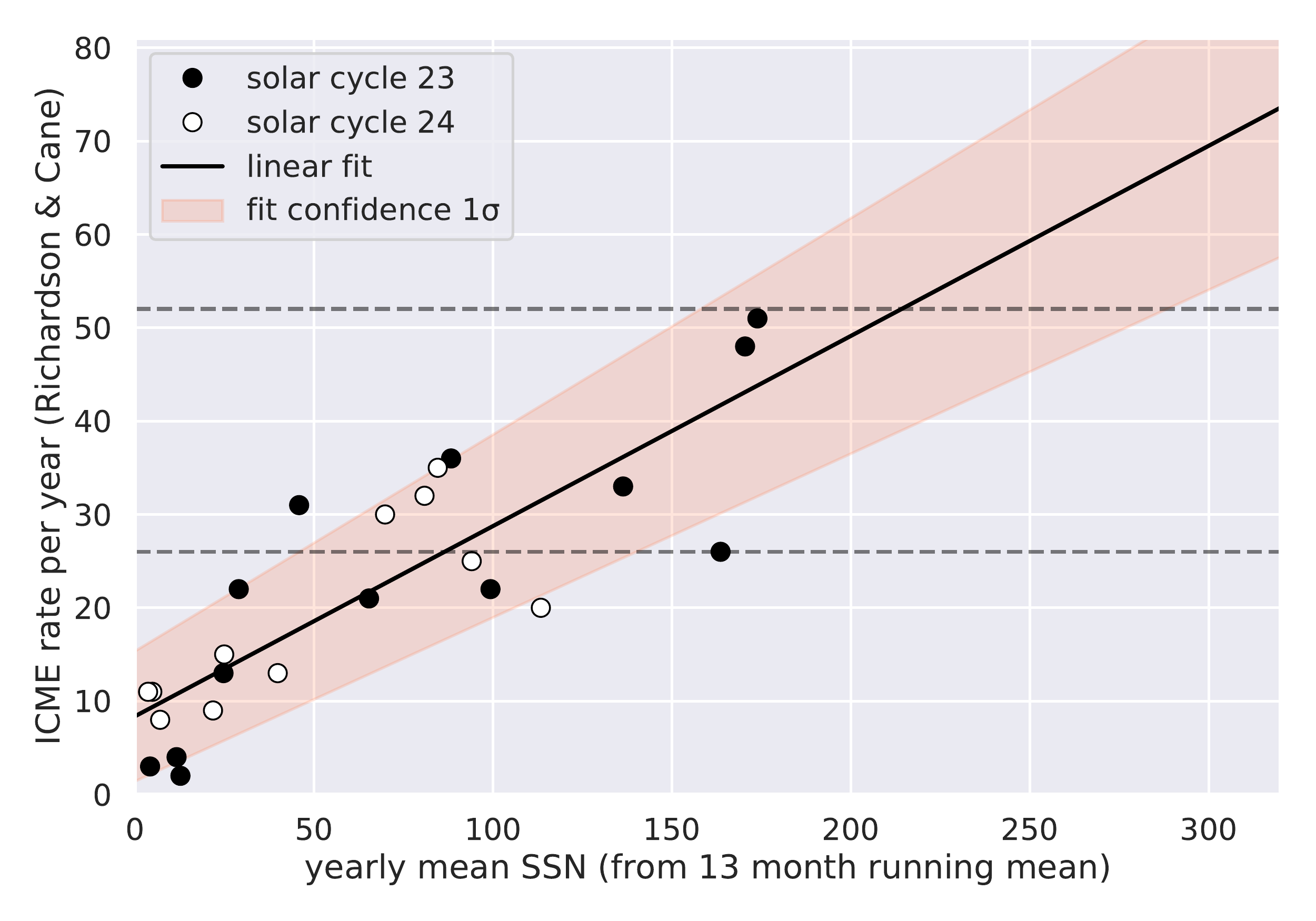}
\caption{Linear fit of yearly averaged sunspot number, from a 13 month running mean (solar cycle 23: black dots, solar cycle 24: white dots) versus the yearly average ICME rate. A $1\sigma$ confidence interval is plotted, corresponding to the error bars in Equation (1). The two dashed horizontal lines indicate ICME rates of 26 and 52 per year, corresponding to one ICME every two weeks or one ICME every week at a heliospheric location, respectively. }
\label{fig:ssn_icr}
\end{figure}

In order to predict the ICR for solar cycle 25 at a position in the inner heliosphere such as the Parker Solar Probe, we need to establish a relationship between a solar cycle progress indicator and the ICR \citep{Kilpua_2011_min_ICMEs,Richardson_2013}. We choose the sunspot number (SSN) as it is the standard parameter for such analyses.

The relationship between CMEs, ICMEs and solar activity indices such as the SSN has been studied extensively in the past \citep{webb1994, gopalswamy2010, li2018, Lamy2019}. In general, the CME rate as measured in coronagraphs follows solar activity indices such as the number and complexity of active regions, reflected by the sunspot number, but \cite{Lamy2019} find the association of CMEs with streamers is the clearest link. However, these associations remain a nuanced problem, and many factors such as progress in the solar cycle, the hemisphere where an eruption originates, coronal magnetic field changes \citep[e.g.][]{gopalswamy2014} and CMEs from stealth events or filament eruptions affect the relationship. In this study we concentrate on a linear fit between the ICME and SSN without considering other influences for a first approximation.

\textbf{Figure \ref{fig:ssn_icr}} shows the yearly mean SSN from SIDC\footnote{\url{http://sidc.be/silso/home}} for solar cycles 23 (SC23) and 24 (SC24), with SC23 defined as the years 1996 to 2008 and SC24 covering 2009 to 2019, for simplicity. This yearly mean was generated from a 13-month running mean of the daily SSN. We plot this against the ICME rate in the \citet{Richardson_2010_ICME_list_online} list\footnote{\url{http://www.srl.caltech.edu/ACE/ASC/DATA/level3/icmetable2.htm}}. 

A linear relationship is fitted as: 

\begin{equation}
ICR =  (0.20 \pm 0.03)  * SSN +  8.4 \pm 7.0, 
\end{equation}
with ICR in a unit of events per year, and SSN given as a yearly average. The error in the slope was taken from a standard procedure (\textit{scipy.stats.linregress} in python) and the error in the intercept term is the average standard deviation between the linear fit and the data points, yielding a $1\sigma$ confidence interval.

With the \citet{Richardson_2010_ICME_list_online} list, we thus find a considerable spread in the yearly ICME rate compared to the fit that arises from the stochastic nature of CME eruptions and ICME impacts. Nevertheless, the Pearson correlation coefficient is 0.84, and it is also clear from physical reasoning that in a simplified picture a higher sunspot number results in more active regions, more CMEs, and more ICMEs. Thus, we consider this relationship robust in representing an approximate connection between the SSN and the ICR.



\section{Results} \label{sec:results}

We now proceed to predict the ICR for solar cycle 25 and calculate the number of ICMEs the Parker Solar Probe spacecraft is expected to encounter for different heliocentric distances until the end of the nominal mission in 2025. We then model the magnetic field signatures of possible double flux rope encounters close to the Sun.

\subsection{ICME rate prediction for solar cycle 25} 

\begin{figure*}[t]
\noindent\includegraphics[width=\linewidth]{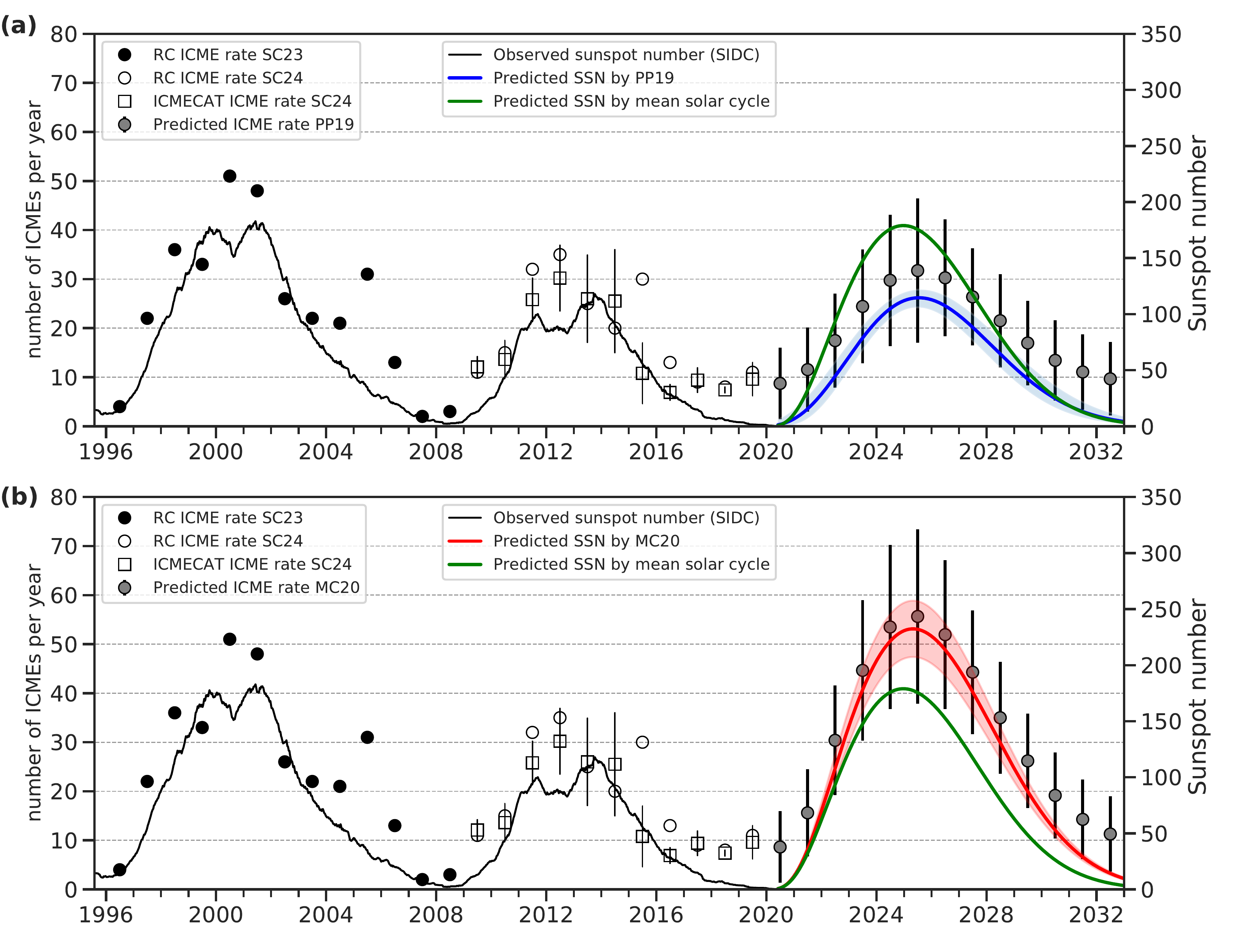}
\caption{Observations and forecasts for the sunspot number and ICME rates. (a) The Richardson and Cane (black and white dots) ICME rate is shown for solar cycle 23/24, as well as the ICME rate in ICMECAT (white squares) including the spread of the ICME rate per year obtained from ICMECAT as  error bars. For solar cycle 25, a Hathaway function for a mean solar cycle is given (green solid line) as well as the predicted SSN by PP19 (blue solid line) including their $1\sigma$ uncertainty. The derived ICME rate is shown as grey dots, including uncertainties as error bars (see text). (b) The MC20 sunspot number prediction, which is clearly above a mean solar cycle model, and the derived ICME rate, is shown in a similar style. Note that the scaling of the axes is similar on both panels. }
\label{fig:icr}
\end{figure*}

Given the linear relationship between the ICR and SSN, we can now proceed to predict the ICR for solar cycle 25. 
Overviews of solar cycle prediction methods are given by \citet{hathaway2015} and \citet{Petrovay_2020}. In 2019, an expert panel consisting of NOAA, NASA and ISES issued a consensus prediction that is published on the NOAA website\footnote{\url{https://www.swpc.noaa.gov/products/solar-cycle-progression}} and available as a JSON file (downloaded in May 2020). The predicted SSN values are available in monthly resolution, and we fit those and another two time series that are given for the lower and upper uncertainty ranges with a \citet{hathaway1994} function of the form:

\begin{equation}
SSN (t) = A\left(\frac{t-t_0}{b}\right)^3 \left[exp\left(\frac{t-t_0}{b}\right)^2    -c \right]^{-1} 
\end{equation}

The fit parameters emerge as $A = 256~[228.2,~273.3]$, $b=60.3~[54.4,~67.4]$, $c=0.35~[0.48,~0.34]$, with the values for the functions with the uncertainty ranges given in brackets (first the lower, then the upper error boundary). Variable $t_0$ is a start time that is returned here by the fit as 2019 September 6. This SSN model peaks at a value of 115 in July 2025. For comparison, the mean maximum SSN for all solar cycles since 1755 is 179, when smoothing the SSN with a 13-month running mean. Thus, the panel forecasts a cycle with a maximum that is a factor 0.64 or about 2/3 of the historic average. This peak would almost exactly match solar cycle 24, which had the fourth lowest maximum SSN since 1755. We call this the panel prediction 2019 (PP19).

\citet{mcintosh2020} introduced a much higher SSN prediction based on timings of terminator events, which signal the end of a solar magnetic cycle \citep{leamon2020}. We call this the MC20 forecast. We created a Hathaway function with parameters $A=444~[396,~492]$, $b=60$, $c=0.8$, including the $1\sigma$ error range given by these authors, which fit the MC20 prediction of a maximum SSN of 232 in April 2025. This maximum SSN is very nearly twice as high as for PP19, and it is a factor 1.3 higher than the historic average. This forecast would clock in as the 7th strongest solar cycle since 1755. The start time $t_0$  here is set as 2020 February 1. Note that a difference in $t_0$ of a few months will not affect our final results for the expected ICME rates at the Parker Solar Probe spacecraft. Given the large discrepancy between the two forecasts, we further work with both and check the results for each predicted ICME rate individually.

\textbf{Figure \ref{fig:icr}a} shows the ICR prediction for solar cycle 25 based on the Hathaway model curve for the SSN given by PP19 and then further derived from a linear relationship between SSN and ICR. For comparison, we plot the average solar cycle peaking with an SSN of 179 as a green solid line. It is seen that the PP19 prediction is clearly below an average solar cycle. Additionally, we show the yearly Richardson and Cane ICR as well as the ICMECAT ICR with a standard deviation of the rate at different in situ locations. \textbf{Figure \ref{fig:icr}b} shows the prediction for solar cycle 25 using the Hathaway function consistent with the MC20 SSN prediction. 

We have also plotted an uncertainty range for the ICR predictions in Figure \ref{fig:icr} based on 3 different sources: (1) the given uncertainties in the SSN prediction, (2) the error in the slope of the linear fit from the SSN and ICR correlation, (3) and the standard deviation from the ICMECAT yearly rate. These errors are added together as squares first, and then the root of their sum is taken as the final error range, as the sources of errors are all independent variables. We see that with all these uncertainties taken into account, the MC20 ICR peak is between 40 and 70 ICMEs per year, which would even exceed the observed rates in SC23, while the PP19 ICR would remain around the low values observed in SC24, peaking between 20 and 40 ICMEs per year. We now figure out how these results affect the observations by the Parker Solar Probe close to the Sun.



\subsection{Implications for Parker Solar Probe close encounters}\label{sec:parker}

\begin{figure*}[t]
\noindent\includegraphics[width=\linewidth]{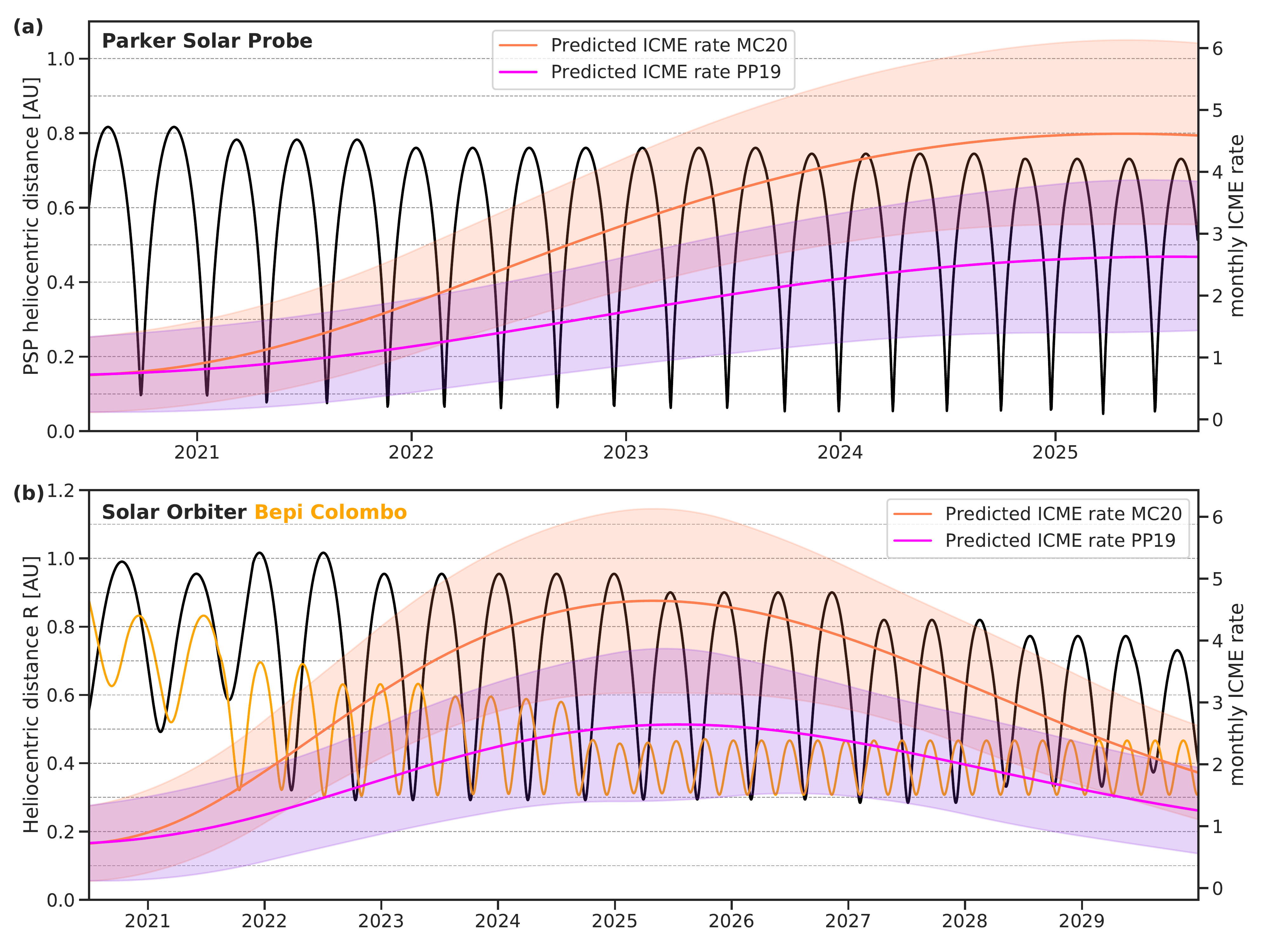}
\caption{Predicted monthly ICME rate for solar cycle 25 (right y-axis) and the heliocentric distance (left y-axis) for (a) Parker Solar Probe, and (b) for Solar Orbiter (black) and Bepi Colombo (orange).  ICME rates based on the PP19 (violet) and MC20 (coral) sunspot number predictions are shown, including an uncertainty range from several causes (see text). The times terminate at the end times of the available spice kernels from each spacecraft, and the position for Bepi Colombo is continued at Mercury after orbit insertion in December 2025.}
\label{fig:prediction}
\end{figure*}

\textbf{Figure~\ref{fig:prediction}} shows the heliocentric distances for PSP, Solar Orbiter and Bepi Colombo and the monthly ICME rate determined from the previous section including error bars. For PSP in \textbf{in Figure~\ref{fig:prediction}a}, this covers the nominal part of the mission, while for Solar Orbiter and Bepi Colombo in \textbf{Figure~\ref{fig:prediction}b}, with the latter to be inserted into orbit around Mercury in December 2025, we extended the plot to the end of 2029. To convert the results of the ICR from a yearly to daily time resolution we used a 2nd order spline fit to interpolate on the yearly values. The figure clearly shows that there is little overlap between the PP19 and MC20 based forecasts for the ICR. 

With a predicted daily ICR for solar cycle 25 in our hands given in Hathaway function format, we can now easily calculate for each day until the end of 2025 how much time PSP spends below a given heliocentric distance and how many ICMEs it is expected to observe there. \textbf{Table 1} gives an overview of the total number of ICMEs expected at PSP below several heliocentric distances for both SSN models. 
These numbers are valid from 2020 July 1 to 2025 August 31, when PSP will have completed 24 solar flybys. For the total mission time since July 2020 and for all heliocentric distances, the PP19 prediction gives $109 \pm 57$ ICME events, whereas the model based on MC20 expects $186 \pm 67$ total ICMEs, which is almost a factor of 2 higher. Below $0.3$~AU, which was never visited by spacecraft before the launch of PSP, the numbers range from 7 to 34, when we include error bars on both the PP19 and MC20 forecasts. The most important result for our purposes is that PSP is expected to observe between 1 and 7 ICMEs below 0.1 AU, which is sufficiently high to consider the special observing geometry of these encounters further in the next section.

\begin{deluxetable}{ccc}\label{tab:psp} 
\tablenum{1}
\tablecaption{\textbf{Total number of ICMEs expected to be observed by PSP in situ}. Each row gives the predicted total number of ICME encounters by PSP below the given heliocentric distance, from 2020 July 1 until 2025 August 31. }
\tablehead{
\colhead{R [AU]} & \colhead{PP19} & \colhead{MC20} } 
\startdata
$all$  & $109 \pm 57$ & $186 \pm 67$ \\
$< 0.3$  & $15 \pm 8$ & $25 \pm 9$ \\
$< 0.2$  & $8 \pm 4$ & $14 \pm 5$ \\
$< 0.1$  & $3 \pm 2$  & $5 \pm 2$ \\
\enddata
\end{deluxetable}


\subsection{Modeling the PSP double observation of ICME flux ropes}\label{sec:3dcore}

\begin{figure*}[t]
\noindent\includegraphics[trim={0 2cm 0 0},width=\linewidth]{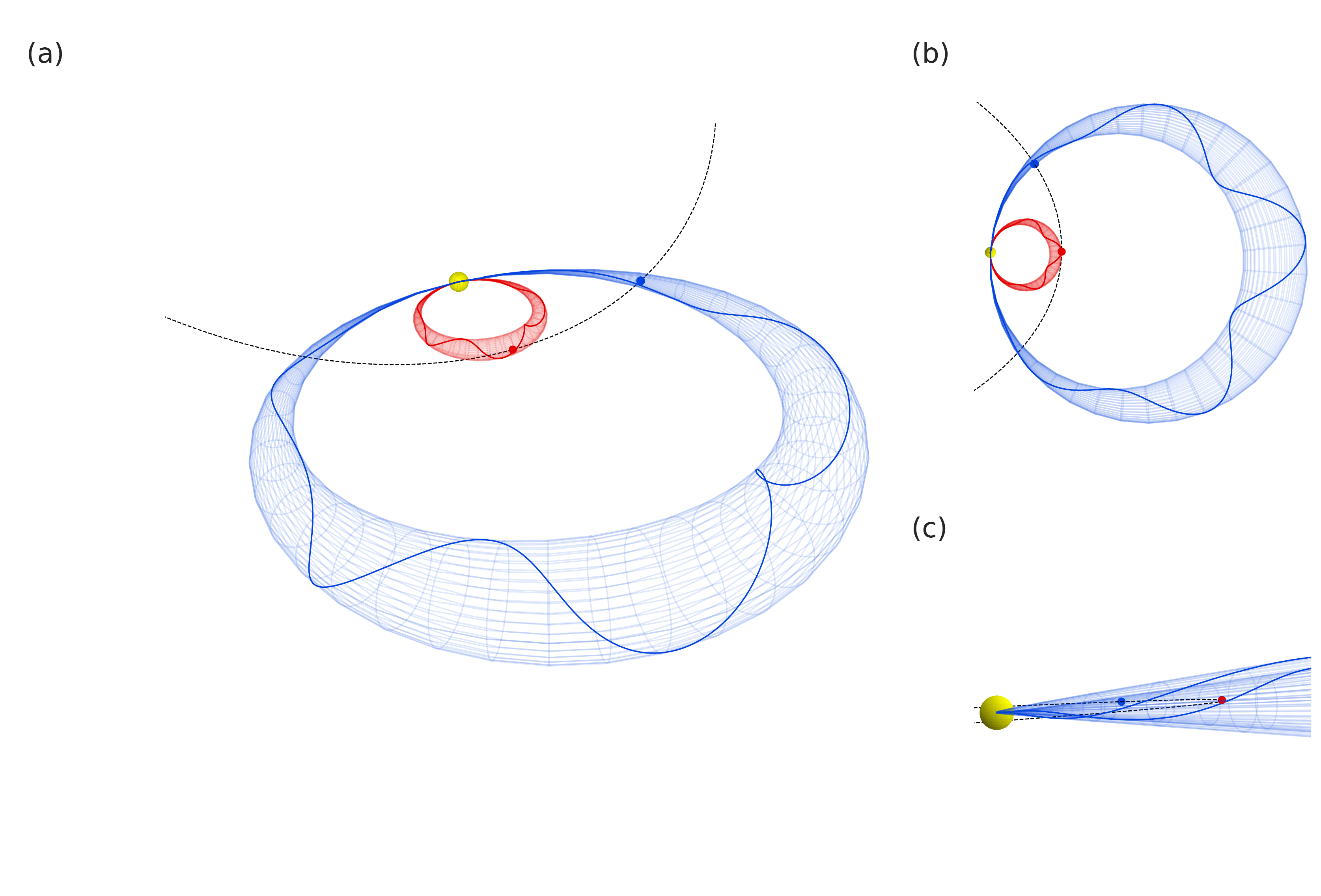}
\caption{Modeling a potential ICME flux rope double crossing event by a virtual Parker Solar Probe with 3DCORE. (a) The 3DCORE flux rope is shown in red at a simulation time of $t_{launch}+4$~hours, when PSP enters the flux rope for the first time, with the PSP trajectory indicated as a dashed line. The blue flux rope represents the simulation time of $t_{launch}+26$~hours, approximately when the virtual PSP spacecraft starts to cross the leg of the flux rope. Panel (b) shows the same model configuration in a view from solar north, and in panel (c) only the blue flux rope in an edge-on view at a 90 degree angle to the CME direction is given, with the position of PSP at simulation time of 4 (26) hours shown as a red (blue) dot. This demonstrates a considerable movement of PSP for an imaging observer during the solar close encounters. This figure is available as an animation in the online version of the article and at \url{https://www.youtube.com/watch?v=VNC2lsw-UtU}.}
\label{fig:double_3D}
\end{figure*}

\begin{figure}[t]
\noindent\includegraphics[width=\linewidth]{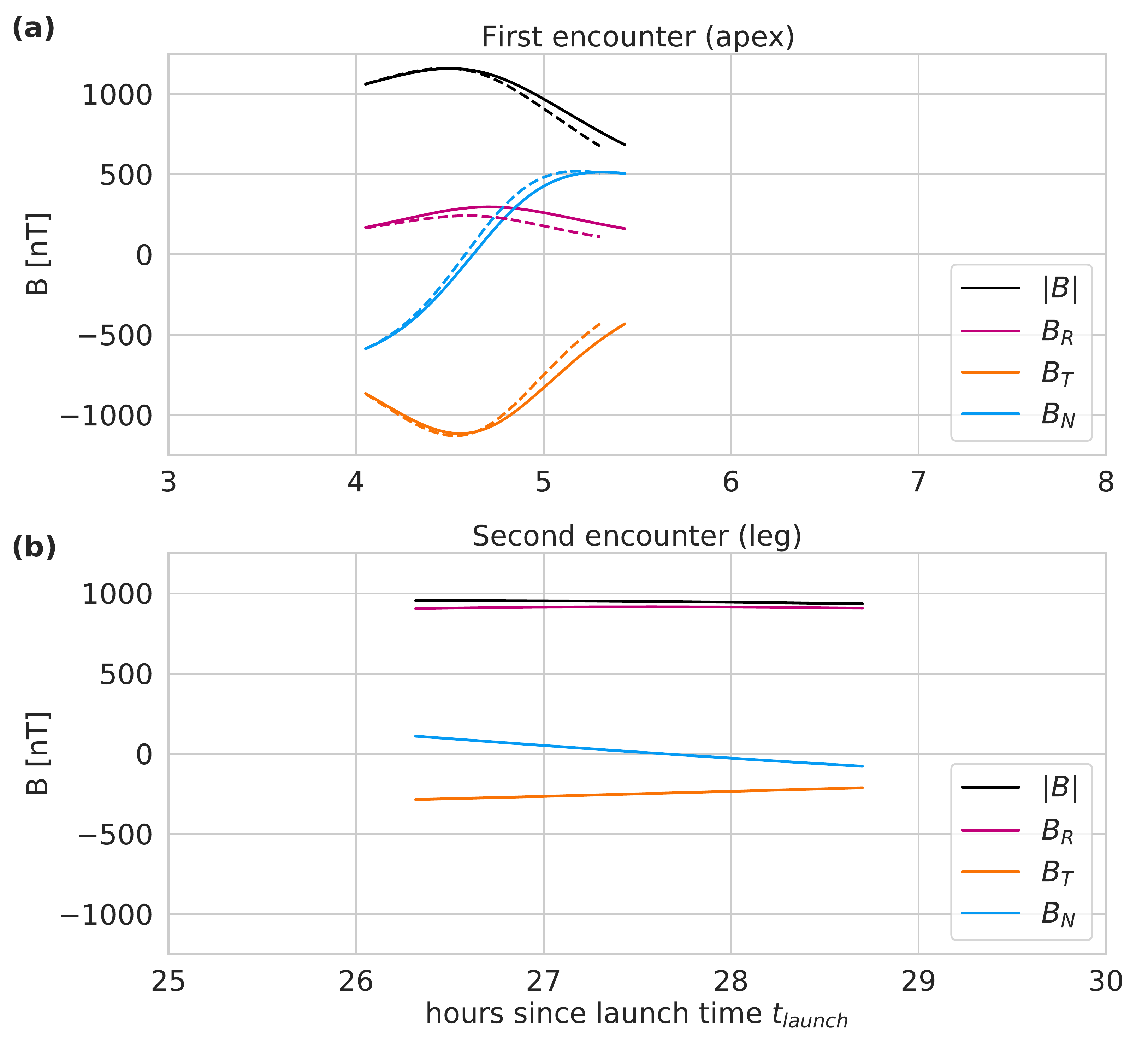}
\caption{(a) Magnetic field components in RTN coordinates as observed by the virtual PSP spacecraft as it moves along its trajectory (solid lines), compared to the stationary case (dashed lines). The time is given in hours since the launch time $t_{launch}$. (b) Similar measurements during the leg encounter, where the time is shifted compared to panel (a) by +~22 hours. Note that the virtual stationary spacecraft only impacts the flux rope apex, and not the leg. The axis ranges are similar in both panels and cover 5 hours of in situ observation. }
\label{fig:double_insitu}
\end{figure}

We now ask ourselves what happens when PSP observes a CME with high relative speed during close solar flybys at $< 0.1$~AU and show a simulated ICME encounter of such a situation. As first presented by \citet{moestl_2018} and recently updated by Weiss et al. (2020, submitted to ApJS), 3DCORE is a semi-empirical 3D flux rope model for the magnetic field inside a CME. It can be used for both fitting available in situ observations and for forward modeling of synthetic magnetic field measurements. It uses the force-free constant twist Gold-Hoyle flux rope model \citep{gold_hoyle_1960} in an approximated 3D configuration to calculate the magnetic field components, which are shown here in radial-tangential-normal RTN coordinates, similar to how PSP magnetic field observations are available. The geometrical shape is described as a tapered torus attached to the Sun, and here we use a circular cross-section, although elliptically flattened cross-sections are available within the model framework too. The flux rope expands according to empirical relationships for the decay of the flux rope magnetic fields, and moves with a drag based modeling approach \citep{vrsnak_2013}.

For the simulated PSP in situ ICME encounter, we assume it happens during its 12th solar flyby on 2022 June 1, reaching at minimum distance to the Sun of 0.061 AU or 13.1 solar radii. Between late 2021 and 2023 there is a series of 7 PSP orbits with the same perihelion distance. We place the flux rope apex at 5 solar radii, from where it launches at the time $t_{launch}=$ 2022 June 1 20:00 UT. The CME direction, defined by the flux rope apex direction, is chosen as such that it points towards the flyby perihelion at 145 degree longitude and 2.5 degree latitude in HAE coordinates. The initial CME speed is 400 km~s$^{-1}$. The flux rope axis inclination is 0 degree to the solar equatorial plane, and it has a left-handed magnetic field chirality, with its axis pointing to solar east. This makes it a south-east-north (SEN) type of magnetic flux rope \citep{bothmer1998}. It propagates into a uniform 400 km~s$^{-1}$ background solar wind, and all other parameters are chosen to represent an average ICME, as it would reach a radial size of 0.24~AU and a maximum total magnetic field strength of 12~nT at 1 AU. 

\textbf{Figure~\ref{fig:double_3D}} gives an overview of the modeling results, visualizing the flux rope shape (red) at $t_{0} + 4$~hours, just as the CME apex impacts PSP. About 1 day later, at about $t_{0} + 26$~h, PSP passes through the model magnetic flux rope (blue) for the 2nd time, this time through the flux rope leg. This means that PSP may essentially make a multipoint in situ CME observation with a single spacecraft, sampling the CME at more than one position within the flux rope and at different times. Clearly, the details of such an encounter would depend on the CME parameters, in particular regarding speed, shape, direction and orientation, with respect to the PSP position and timing.  We further want to illustrate what is expected to happen in an idealized situation.
 
In \textbf{Figure~\ref{fig:double_insitu}a}, the signature (solid lines) of a virtual spacecraft that observes the CME along the real PSP trajectory is seen, exhibiting a typical flux rope signature of an expanding low-inclination flux rope of the SEN type, with $B_T$ making a unipolar excursion to the $-T$ direction (solar east) and $B_N$ reversing its sign from south to north, while $B_R$ stays near zero. The asymmetry in the total field arises from the flux rope expansion during the measurement  as well as the aging of the flux rope as it passes over PSP \citep{Farrugia_1993_expansion}, and other physical processes such as deflection and reconnection with the surrounding solar wind are neglected here.  Due to the small heliocentric distance, where the CME flux rope has not yet expanded much, the observation only takes around 1.5 hours, and peaks at a total field magnitude of roughly $10^3$~nT. PSP moves here at a maximum speed of 161 km~s$^{-1}$ during perihelion. We compare this to a virtual PSP spacecraft that observes the CME magnetic field while staying still at the CME apex position in \textbf{Figure \ref{fig:double_insitu}a}, shown as dashed lines. There is only very little difference between the moving and stationary cases, from which no additional information on the magnetic field structure of the flux rope can be gained.

However, about 26 hours after the CME launch, the virtual spacecraft intercepts the CME flux rope a second time, and in \textbf{Figure~\ref{fig:double_insitu}b} we see that now the $B_N$ component, though staying near zero, is essentially reversed compared to the first encounter, as the virtual PSP now observes the flux rope from the inside to the outside - compared to an outside-to-inside trajectory during the apex encounter. The radial $B_R$ component is strongly elevated and almost similar to the total field as the axial field of the flux rope points away from the Sun at the western leg. The sign of the $B_T$ component is similar to the first encounter, but the magnitude is strongly decreased. This observation lasts for about 2 hours in the simulation, as during one day the flux rope has already somewhat expanded, even though the flux rope cross-section diameter narrows from the apex to the leg in the 3DCORE model.  Without modeling such as shown here the connection between two such possible observations as a double encounter of the same flux rope would likely not be made.

Simultaneous imaging of a double-crossing event by coronagraphs and heliospheric imagers onboard SOHO, STEREO-A, PSP or Solar Orbiter could further substantiate the interpretation that PSP indeed crosses a CME flux rope twice. When multiple viewpoints are available, forward modelling techniques \citep[e.g.][]{thernisien_2006} provide results on CME direction, speed and longitudinal extent. They should help in discriminating whether the two in situ signatures proposed in Figure~\ref{fig:double_insitu} may arise from the same CME flux rope or if rather two flux ropes from different CMEs are observed. This becomes more important later in the PSP mission as the spacecraft perihelia become smaller while the CME rate increases (see Figure~\ref{fig:prediction}), likely leading to multiple CMEs erupting in quick succession during solar maximum \citep[e.g.][]{schrijver2011}.

The very different magnetic signatures that could be observed during both PSP flux rope encounters would be an extremely valuable hint to constrain the 3D CME magnetic field structure, and may help in clarifying pressing questions on the origins of CMEs. On the other hand, a lack of the 2nd encounter of the CME flux rope, where the CME was clearly observed in coronagraphs to have the morphology of a flux rope with its symmetry axis close to the equator \citep[e.g.][]{thernisien_2006,vourlidas2013, wood2017}, would raise questions on the general validity of the flux rope picture, and how the magnetic field lines at the flux rope leg close in on the corona and whether they are already detached from the Sun. However, from bi-directional electron streaming observations, it is expected that for such small heliocentric distances only very few magnetic field lines would have already separated from the Sun \citep[e.g.][]{shodhan2000,nieves_chinchilla_2020}.  A significant deviation of the measurements with the prediction from 3DCORE would indicate that additional physical processes not captured by this version of the model are central to the coronal evolution of CMEs, for example a deflection in direction, a different expansion law or magnetic reconnection with the background solar wind.

\section{Conclusions}

We have shown that double crossings of ICME flux ropes during close solar flybys of the Parker Solar Probe (PSP) spacecraft are possible. To this end, we have made an update of the largest available database of interplanetary coronal mass ejections (ICMEs), with 739 events cataloged since 2007. Together with the ICME list by \citet{Richardson_2010_ICME_list_online}, which includes all ICMEs measured in the solar wind near Earth since 1996, we have made a linear fit as an approximate relationship of the yearly ICME rate with the yearly, smoothed sunspot number as a proxy for solar activity, including an uncertainty arising from the stochastic nature of ICME observations that results in the variation of the yearly ICME rate for different places in the inner heliosphere.

We have then forecasted the ICME rate for the next solar cycle 25 based on two different predictions for the sunspot number, and determined that between 1 and 7 CMEs are expected to be observed by PSP at distances $< 0.1$~AU to the Sun. We then modeled the consequences for possible in situ observations by the PSP during a close solar flyby by combining our semi-empirical flux rope model 3DCORE with the predicted PSP flight trajectory in an idealized setting. The simulated in situ measurements show that although PSP crosses the flux rope at high relative speed, the observation of the magnetic field components near the flux rope apex stays almost identical as compared to the measurements made by a stationary spacecraft. However, for the modeled flux rope with a low inclination to the solar equatorial plane, PSP crosses the leg about a day later, which in our simulation gives a telltale reversal in the sign of the $B_N$ component and a strongly increased $B_R$ field compared to the apex encounter. In reality, we expect that a higher inclination  of the flux rope will decrease in the probability for PSP to observe a double crossing, while a flatter cross section would increase its chances, but a parameter study to look in detail at the influence of different parameter settings goes beyond the scope of the current paper. 
Additionally, imaging data from PSP itself, SOHO, STEREO-A or Solar Orbiter may provide context to further prove that indeed the same CME flux rope was crossed by PSP twice.

In summary, we think these are highly interesting results which could strongly constrain models for predicting the magnetic field components of CMEs, an unsolved problem in space weather prediction, and could even shed light on the early stages of CME eruption and formation \citep[see also][]{alhaddad2019}. Future studies may look into more complex physics surrounding the eruption process of the CME such as channeling \citep{Moestl_2015} or deflection \citep[e.g.][]{gopalswamy2009a}, and the interaction of the flux rope with the background solar wind \citep[e.g.][]{ruffenach2015}, to see whether there are unexpected features observed concerning magnetic reconnection or ICME deformations during the putative double-crossing events. The results presented in this work therefore give an indicator of the type of measurements combined with simulations that we hope to make in the upcoming years. 

It has not escaped our notice that the \citet{mcintosh2020} based prediction for the ICME rate would lead to the the highest rates of ICME impacts at Earth in 40 years, which, if true, will make mitigation of destructive effects of space weather on technological assets increasingly urgent for humankind.

\acknowledgments

C.M., A.J.W, R.L.B, M.A., T.A., M.B and J.H. thank the Austrian Science Fund (FWF): P31265-N27, P31521-N27, P31659-N27. D.S. is supported by STFC consolidated grant number ST/5000240/1. N.L. acknowledges NASA grant 80NSSC20K0700. Sunspot numbers were thankfully provided by SILSO Royal Observatory of Belgium, Brussels (\url{http://sidc.be/silso/home}). 

This research used version 1.1.0 (doi: 10.5281/zenodo.3604353) of the SunPy open source software package \citep{sunpy2020}. This research made use of Astropy (\url{http://www.astropy.org}) a community-developed core Python package for Astronomy \citep{astropy_2013, astropy_2018}. This research made use of HelioPy, a community-developed Python package for space physics [doi:10.5281/zenodo.3739113]. 

The version of the ICMECAT catalog used in this paper is permanently available at \url{https://doi.org/10.6084/m9.figshare.6356420.v3}. The code to produce the results in this paper is accessible in two jupyter notebooks at \url{https://github.com/helioforecast/Papers/tree/master/Moestl2020_PSP_rate} and the code, figures and animations for this paper are additionally archived at \url{https://doi.org/10.6084/m9.figshare.12563765}.











\bibliography{chrisbib}

\begin{thebibliography}{}
\expandafter\ifx\csname natexlab\endcsname\relax\def\natexlab#1{#1}\fi
\providecommand{\url}[1]{\href{#1}{#1}}

\bibitem[{{Al-Haddad} {et~al.}(2019){Al-Haddad}, {Lugaz}, {Poedts}, {Farrugia},
  {Nieves-Chinchilla}, \& {Roussev}}]{alhaddad2019}
{Al-Haddad}, N., {Lugaz}, N., {Poedts}, S., {et~al.} 2019, \apj, 884, 179

\bibitem[{{Astropy Collaboration} {et~al.}(2013){Astropy Collaboration},
  {Robitaille}, {Tollerud}, {Greenfield}, {Droettboom}, {Bray}, {Aldcroft},
  {Davis}, {Ginsburg}, {Price-Whelan}, {Kerzendorf}, {Conley}, {Crighton},
  {Barbary}, {Muna}, {Ferguson}, {Grollier}, {Parikh}, {Nair}, {Unther},
  {Deil}, {Woillez}, {Conseil}, {Kramer}, {Turner}, {Singer}, {Fox}, {Weaver},
  {Zabalza}, {Edwards}, {Azalee Bostroem}, {Burke}, {Casey}, {Crawford},
  {Dencheva}, {Ely}, {Jenness}, {Labrie}, {Lim}, {Pierfederici}, {Pontzen},
  {Ptak}, {Refsdal}, {Servillat}, \& {Streicher}}]{astropy_2013}
{Astropy Collaboration}, {Robitaille}, T.~P., {Tollerud}, E.~J., {et~al.} 2013,
  \aap, 558, A33

\bibitem[{{Bothmer} \& {Schwenn}(1998)}]{bothmer1998}
{Bothmer}, V., \& {Schwenn}, R. 1998, \angeo, 16, 1

\bibitem[{{Carrington}(1859)}]{carrington1859}
{Carrington}, R.~C. 1859, \mnras, 20, 13

\bibitem[{{Farrugia} {et~al.}(1993){Farrugia}, {Burlaga}, {Osherovich},
  {Richardson}, {Freeman}, {Lepping}, \& {Lazarus}}]{Farrugia_1993_expansion}
{Farrugia}, C.~J., {Burlaga}, L.~F., {Osherovich}, V.~A., {et~al.} 1993, \jgr,
  98, 7621

\bibitem[{{Fox} {et~al.}(2016){Fox}, {Velli}, {Bale}, {Decker}, {Driesman},
  {Howard}, {Kasper}, {Kinnison}, {Kusterer}, {Lario}, {Lockwood}, {McComas},
  {Raouafi}, \& {Szabo}}]{fox2016_psp}
{Fox}, N.~J., {Velli}, M.~C., {Bale}, S.~D., {et~al.} 2016, \ssr, 204, 7

\bibitem[{{Gold} \& {Hoyle}(1960)}]{gold_hoyle_1960}
{Gold}, T., \& {Hoyle}, F. 1960, \mnras, 120, 89

\bibitem[{Gopalswamy {et~al.}(2010)Gopalswamy, Akiyama, Yashiro, \&
  M{\"a}kel{\"a}}]{gopalswamy2010}
Gopalswamy, N., Akiyama, S., Yashiro, S., \& M{\"a}kel{\"a}, P. 2010, in
  Magnetic Coupling between the Interior and Atmosphere of the Sun (Springer),
  289--307

\bibitem[{{Gopalswamy} {et~al.}(2014){Gopalswamy}, {Akiyama}, {Yashiro}, {Xie},
  {M{\"a}kel{\"a}}, \& {Michalek}}]{gopalswamy2014}
{Gopalswamy}, N., {Akiyama}, S., {Yashiro}, S., {et~al.} 2014, \grl, 41, 2673

\bibitem[{{Gopalswamy} {et~al.}(2009){Gopalswamy}, {M{\"a}kel{\"a}}, {Xie},
  {Akiyama}, \& {Yashiro}}]{gopalswamy2009a}
{Gopalswamy}, N., {M{\"a}kel{\"a}}, P., {Xie}, H., {Akiyama}, S., \& {Yashiro},
  S. 2009, \jgr, 114, A00A22

\bibitem[{{Hathaway}(2015)}]{hathaway2015}
{Hathaway}, D.~H. 2015, \lrsp, 12, 4

\bibitem[{{Hathaway} {et~al.}(1994){Hathaway}, {Wilson}, \&
  {Reichmann}}]{hathaway1994}
{Hathaway}, D.~H., {Wilson}, R.~M., \& {Reichmann}, E.~J. 1994, \solphys, 151,
  177

\bibitem[{Jian {et~al.}(2006)Jian, Russell, Luhmann, \& Skoug}]{Jian2006}
Jian, L., Russell, C., Luhmann, J., \& Skoug, R. 2006, Solar Physics, 239, 337

\bibitem[{{Kilpua} {et~al.}(2011){Kilpua}, {Lee}, {Luhmann}, \&
  {Li}}]{Kilpua_2011_min_ICMEs}
{Kilpua}, E.~K.~J., {Lee}, C.~O., {Luhmann}, J.~G., \& {Li}, Y. 2011, Annales
  Geophysicae, 29, 1455

\bibitem[{Lamy {et~al.}(2019)Lamy, Floyd, Boclet, Wojak, Gilardy, \&
  Barlyaeva}]{Lamy2019}
Lamy, P., Floyd, O., Boclet, B., {et~al.} 2019, Space Science Reviews, 215, 39

\bibitem[{{Leamon} {et~al.}(2020){Leamon}, {McIntosh}, {Chapman}, \&
  {Watkins}}]{leamon2020}
{Leamon}, R.~J., {McIntosh}, S.~W., {Chapman}, S.~C., \& {Watkins}, N.~W. 2020,
  \solphys, 295, 36

\bibitem[{{Li} {et~al.}(2018){Li}, {Luhmann}, \& {Lynch}}]{li2018}
{Li}, Y., {Luhmann}, J.~G., \& {Lynch}, B.~J. 2018, \solphys, 293, 135

\bibitem[{{Love} {et~al.}(2019){Love}, {Hayakawa}, \&
  {Cliver}}]{love2019_may21}
{Love}, J.~J., {Hayakawa}, H., \& {Cliver}, E.~W. 2019, Space Weather, 17, 1281

\bibitem[{{Love} {et~al.}(2015){Love}, {Rigler}, {Pulkkinen}, \&
  {Riley}}]{love2015_historic}
{Love}, J.~J., {Rigler}, E.~J., {Pulkkinen}, A., \& {Riley}, P. 2015, \grl, 42,
  6544

\bibitem[{{McIntosh} {et~al.}(2020){McIntosh}, {Chapman}, {Leamon}, {Egeland},
  \& {Watkins}}]{mcintosh2020}
{McIntosh}, S.~W., {Chapman}, S.~C., {Leamon}, R.~J., {Egeland}, R., \&
  {Watkins}, N.~W. 2020, arXiv e-prints, arXiv:2006.15263

\bibitem[{{M{\"o}stl} {et~al.}(2015){M{\"o}stl}, {Rollett}, {Frahm}, {Liu},
  {Long}, {Colaninno}, {Reiss}, {Temmer}, {Farrugia}, {Posner}, {Dumbovi{\'c}},
  {Janvier}, {D{\'e}moulin}, {Boakes}, {Devos}, {Kraaikamp}, {Mays}, \& {Vr{\v
  s}nak}}]{Moestl_2015}
{M{\"o}stl}, C., {Rollett}, T., {Frahm}, R.~A., {et~al.} 2015, Nature
  Communications, 6, 7135

\bibitem[{M{\"o}stl {et~al.}(2017)M{\"o}stl, Isavnin, Boakes, Kilpua, Davies,
  Harrison, Barnes, Krupar, Eastwood, Good, Forsyth, Bothmer, Reiss,
  Amerstorfer, Winslow, Anderson, Philpott, Rodriguez, Rouillard, Gallagher,
  Nieves-Chinchilla, \& Zhang}]{Moestl2017}
M{\"o}stl, C., Isavnin, A., Boakes, P.~D., {et~al.} 2017, Space Weather, 15,
  955

\bibitem[{{M{\"o}stl} {et~al.}(2018){M{\"o}stl}, {Amerstorfer}, {Palmerio},
  {Isavnin}, {Farrugia}, {Lowder}, {Winslow}, {Donnerer}, {Kilpua}, \&
  {Boakes}}]{moestl_2018}
{M{\"o}stl}, C., {Amerstorfer}, T., {Palmerio}, E., {et~al.} 2018, Space
  Weather, 16, 216

\bibitem[{{M{\"u}ller} {et~al.}(2013){M{\"u}ller}, {Marsden}, {St. Cyr}, \&
  {Gilbert}}]{Mueller_2013}
{M{\"u}ller}, D., {Marsden}, R.~G., {St. Cyr}, O.~C., \& {Gilbert}, H.~R. 2013,
  \solphys, 285, 25

\bibitem[{{Nieves-Chinchilla} {et~al.}(2020){Nieves-Chinchilla}, {Szabo},
  {Korreck}, {Alzate}, {Balmaceda}, {Lavraud}, {Paulson}, {Narock}, {Wallace},
  {Jian}, {Luhmann}, {Morgan}, {Higginson}, {Arge}, {Bale}, {Case}, {Wit},
  {Giacalone}, {Goetz}, {Harvey}, {Jones-Melosky}, {Kasper}, {Larson}, {Livi},
  {McComas}, {MacDowall}, {Malaspina}, {Pulupa}, {Raouafi}, {Schwadron},
  {Stevens}, \& {Whittlesey}}]{nieves_chinchilla_2020}
{Nieves-Chinchilla}, T., {Szabo}, A., {Korreck}, K.~E., {et~al.} 2020, \apjs,
  246, 63

\bibitem[{{Petrovay}(2020)}]{Petrovay_2020}
{Petrovay}, K. 2020, Living Reviews in Solar Physics, 17, 2

\bibitem[{{Price-Whelan} {et~al.}(2018){Price-Whelan}, {Sip{\H{o}}cz},
  {G{\"u}nther}, {Lim}, {Crawford}, {Conseil}, {Shupe}, {Craig}, {Dencheva},
  {Ginsburg}, {VanderPlas}, {Bradley}, {P{\'e}rez-Su{\'a}rez}, {de Val-Borro},
  {Paper Contributors}, {Aldcroft}, {Cruz}, {Robitaille}, {Tollerud},
  {Coordination Committee}, {Ardelean}, {Babej}, {Bach}, {Bachetti}, {Bakanov},
  {Bamford}, {Barentsen}, {Barmby}, {Baumbach}, {Berry}, {Biscani}, {Boquien},
  {Bostroem}, {Bouma}, {Brammer}, {Bray}, {Breytenbach}, {Buddelmeijer},
  {Burke}, {Calderone}, {Cano Rodr{\'\i}guez}, {Cara}, {Cardoso}, {Cheedella},
  {Copin}, {Corrales}, {Crichton}, {D{\textquoteright}Avella}, {Deil},
  {Depagne}, {Dietrich}, {Donath}, {Droettboom}, {Earl}, {Erben}, {Fabbro},
  {Ferreira}, {Finethy}, {Fox}, {Garrison}, {Gibbons}, {Goldstein}, {Gommers},
  {Greco}, {Greenfield}, {Groener}, {Grollier}, {Hagen}, {Hirst}, {Homeier},
  {Horton}, {Hosseinzadeh}, {Hu}, {Hunkeler}, {Ivezi{\'c}}, {Jain}, {Jenness},
  {Kanarek}, {Kendrew}, {Kern}, {Kerzendorf}, {Khvalko}, {King}, {Kirkby},
  {Kulkarni}, {Kumar}, {Lee}, {Lenz}, {Littlefair}, {Ma}, {Macleod},
  {Mastropietro}, {McCully}, {Montagnac}, {Morris}, {Mueller}, {Mumford},
  {Muna}, {Murphy}, {Nelson}, {Nguyen}, {Ninan}, {N{\"o}the}, {Ogaz}, {Oh},
  {Parejko}, {Parley}, {Pascual}, {Patil}, {Patil}, {Plunkett}, {Prochaska},
  {Rastogi}, {Reddy Janga}, {Sabater}, {Sakurikar}, {Seifert}, {Sherbert},
  {Sherwood-Taylor}, {Shih}, {Sick}, {Silbiger}, {Singanamalla}, {Singer},
  {Sladen}, {Sooley}, {Sornarajah}, {Streicher}, {Teuben}, {Thomas},
  {Tremblay}, {Turner}, {Terr{\'o}n}, {van Kerkwijk}, {de la Vega}, {Watkins},
  {Weaver}, {Whitmore}, {Woillez}, {Zabalza}, \& {Contributors}}]{astropy_2018}
{Price-Whelan}, A.~M., {Sip{\H{o}}cz}, B.~M., {G{\"u}nther}, H.~M., {et~al.}
  2018, \aj, 156, 123

\bibitem[{{Richardson}(2013)}]{Richardson_2013}
{Richardson}, I.~G. 2013, Journal of Space Weather and Space Climate, 3, A08

\bibitem[{{Richardson} \& {Cane}(2010)}]{Richardson_2010_ICME_list_online}
{Richardson}, I.~G., \& {Cane}, H.~V. 2010, \solphys, 264, 189

\bibitem[{Richardson {et~al.}(2000)Richardson, Cliver, \&
  Cane}]{Richardson2000}
Richardson, I.~G., Cliver, E.~W., \& Cane, H.~V. 2000, Journal of Geophysical
  Research: Space Physics, 105, 18203

\bibitem[{Richardson {et~al.}(2001)Richardson, Cliver, \&
  Cane}]{Richardson2001}
---. 2001, Geophysical Research Letters, 28, 2569

\bibitem[{{Rouillard}(2011)}]{rouillard2011}
{Rouillard}, A.~P. 2011, Journal of Atmospheric and Solar-Terrestrial Physics,
  73, 1201

\bibitem[{{Ruffenach} {et~al.}(2015){Ruffenach}, {Lavraud}, {Farrugia},
  {D{\'e}moulin}, {Dasso}, {Owens}, {Sauvaud}, {Rouillard}, {Lynnyk},
  {Foullon}, {Savani}, {Luhmann}, \& {Galvin}}]{ruffenach2015}
{Ruffenach}, A., {Lavraud}, B., {Farrugia}, C.~J., {et~al.} 2015, Journal of
  Geophysical Research (Space Physics), 120, 43

\bibitem[{{Schrijver} \& {Title}(2011)}]{schrijver2011}
{Schrijver}, C.~J., \& {Title}, A.~M. 2011, Journal of Geophysical Research
  (Space Physics), 116, A04108

\bibitem[{{Shodhan} {et~al.}(2000){Shodhan}, {Crooker}, {Kahler},
  {Fitzenreiter}, {Larson}, {Lepping}, {Siscoe}, \& {Gosling}}]{shodhan2000}
{Shodhan}, S., {Crooker}, N.~U., {Kahler}, S.~W., {et~al.} 2000, \jgr, 105,
  27261

\bibitem[{{SunPy Community} {et~al.}(2020){SunPy Community}, {Barnes}, {Bobra},
  {Christe}, {Freij}, {Hayes}, {Ireland }, {Mumford}, {Perez-Suarez}, {Ryan},
  {Shih}, {Chanda}, {Glogowski}, {Hewett}, {Hughitt}, {Hill}, {Hiware},
  {Inglis}, {Kirk}, {Konge}, {Mason}, {Maloney}, {Murray}, {Panda}, {Park},
  {Pereira}, {Reardon}, {Savage}, {Sip{\H{o}}cz}, {Stansby}, {Jain}, {Taylor},
  {Yadav}, {Rajul}, \& {Dang}}]{sunpy2020}
{SunPy Community}, {Barnes}, W.~T., {Bobra}, M.~G., {et~al.} 2020, \apj, 890,
  68

\bibitem[{{Thernisien} {et~al.}(2006){Thernisien}, {Howard}, \&
  {Vourlidas}}]{thernisien_2006}
{Thernisien}, A.~F.~R., {Howard}, R.~A., \& {Vourlidas}, A. 2006, \apj, 652,
  763

\bibitem[{{Vourlidas} {et~al.}(2013){Vourlidas}, {Lynch}, {Howard}, \&
  {Li}}]{vourlidas2013}
{Vourlidas}, A., {Lynch}, B.~J., {Howard}, R.~A., \& {Li}, Y. 2013, \solphys,
  284, 179

\bibitem[{{Vr{\v{s}}nak} {et~al.}(2013){Vr{\v{s}}nak}, {{\v{Z}}ic}, {Vrbanec},
  {Temmer}, {Rollett}, {M{\"o}stl}, {Veronig}, {{\v{C}}alogovi{\'c}},
  {Dumbovi{\'c}}, {Luli{\'c}}, {Moon}, \& {Shanmugaraju}}]{vrsnak_2013}
{Vr{\v{s}}nak}, B., {{\v{Z}}ic}, T., {Vrbanec}, D., {et~al.} 2013, \solphys,
  285, 295

\bibitem[{{Webb} \& {Howard}(1994)}]{webb1994}
{Webb}, D.~F., \& {Howard}, R.~A. 1994, \jgr, 99, 4201

\bibitem[{{Webb} \& {Howard}(2012)}]{webb2012_review}
{Webb}, D.~F., \& {Howard}, T.~A. 2012, Living Reviews in Solar Physics, 9, 3

\bibitem[{{Weiss} {et~al.}(2020){Weiss}, {M{\"o}stl}, {Amerstorfer}, {Bailey},
  {Reiss}, {Hinterreiter}, {Amerstorfer}, \& {Bauer}}]{weiss2020}
{Weiss}, A.~J., {M{\"o}stl}, C., {Amerstorfer}, T., {et~al.} 2020, arXiv
  e-prints, arXiv:2009.00327

\bibitem[{{Wood} {et~al.}(2017){Wood}, {Wu}, {Lepping}, {Nieves-Chinchilla},
  {Howard}, {Linton}, \& {Socker}}]{wood2017}
{Wood}, B.~E., {Wu}, C.-C., {Lepping}, R.~P., {et~al.} 2017, \apjs, 229, 29

\bibitem[{{Yashiro} {et~al.}(2004){Yashiro}, {Gopalswamy}, {Michalek},
  {St.~Cyr}, {Plunkett}, {Rich}, \& {Howard}}]{yashiro2004}
{Yashiro}, S., {Gopalswamy}, N., {Michalek}, G., {et~al.} 2004, \jgr, 109,
  A07105

\bibitem[{{Zhang} {et~al.}(2007){Zhang}, {Richardson}, {Webb}, {Gopalswamy},
  {Huttunen}, {Kasper}, {Nitta}, {Poomvises}, {Thompson}, {Wu}, {Yashiro}, \&
  {Zhukov}}]{zhang2007}
{Zhang}, J., {Richardson}, I.~G., {Webb}, D.~F., {et~al.} 2007, Journal of
  Geophysical Research (Space Physics), 112, A10102

\end{thebibliography}

\end{document}